\documentstyle[12pt,twoside,fleqn,espcrc1,epsfig]{article}
\title{ Quasielastic $K^{+}$ scattering in nuclei\thanks{This work has been
supported by the program of Human Capital and Mobility of the EU, 
contract n. CHRX--CT 93--0323} }

\author{
A. De Pace \\
\hfill \\
Istituto Nazionale Fisica Nucleare, Sezione di Torino, 
  via Giuria 1, I-10125 Torino, Italy}

\begin{document}
\maketitle

\section*{}

Quasielastic (QE) studies at intermediate energies are an important tool 
to study both nucleonic and nuclear physics issues. 
In particular, the reasons to consider 
$K^{+}$-nucleus scattering have been twofold. On the one hand, the elementary 
$K^{+}N$ cross section is relatively small compared to other hadronic probes,
thus allowing the kaons to penetrate deeper inside the nucleus, making them more
suitable to study collective effects. Furthermore, since the $K^{+}N$ cross
section is dominated by the scalar-isoscalar channel, kaons turn out to be a
quasi-pure probe of this mode. On the other hand, the excess of cross section,
with respect to multiple scattering theory predictions, that has been found in
$K^{+}$-nucleus elastic scattering experiments is still unexplained and may be
interpreted in terms of an enhancement of the in-medium $K^{+}N$ cross section,
$\sigma_{K^{+}N}$. This finding has naturally raised the issue of what might 
be the consequences for QE scattering.

\begin{figure}[t]
\begin{minipage}[t]{78mm}
\epsfig{file=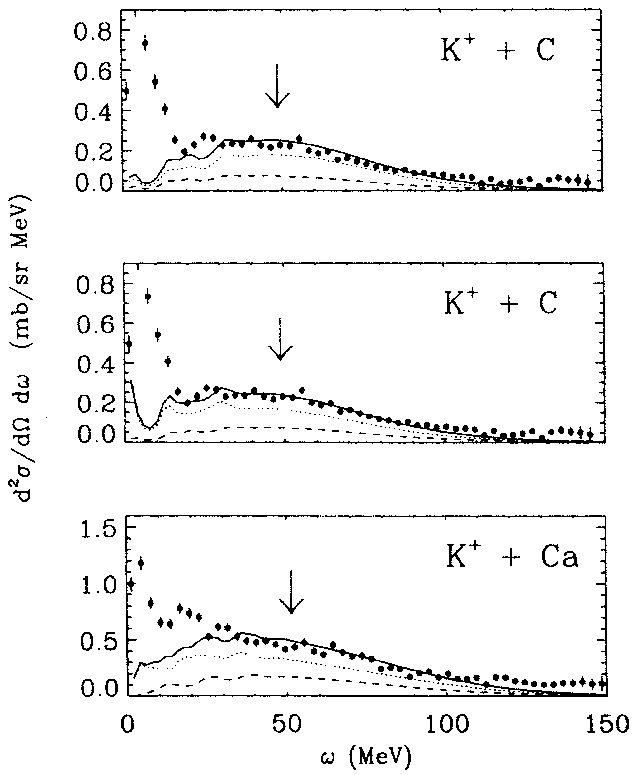,height=9.5cm}
\caption{ QE cross sections for C and Ca at $q=300$ MeV/c from
Ref.\protect\cite{Kor93}. The solid lines are the sum of isoscalar (dot) and
isovector (dash) responses in a finite nucleus relativistic calculation:
Hartree (top), Hartree-RPA (middle and bottom).
}
\label{fig:exp1}
\end{minipage}
\hspace{\fill}
\begin{minipage}[t]{78mm}
\epsfig{file=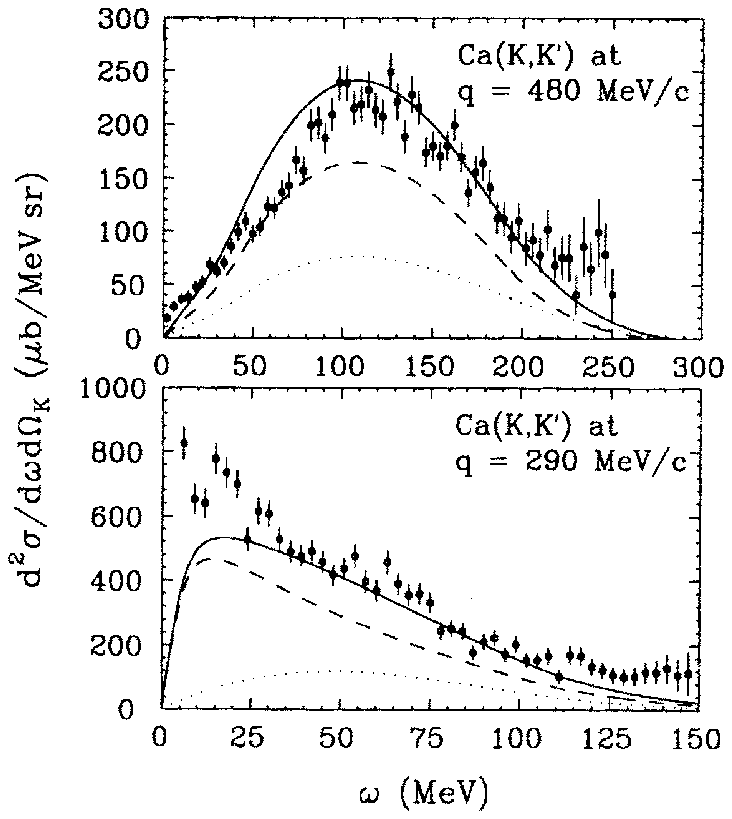,height=8.7cm}
\caption{ QE cross sections for Ca at $q=290$ and 480 MeV/c from
Ref.\protect\cite{Kor95}. The solid lines are the sum of isoscalar (dash) and
isovector (dot) responses in a relativistic Hartree-RPA local density  
calculation.
}
\label{fig:exp2}
\end{minipage}
\end{figure}

A few results from the experiment performed at BNL, taken from 
Refs.\cite{Kor93,Kor95}, are displayed in Figures \ref{fig:exp1} and 
\ref{fig:exp2}. In those papers the data have been compared to calculations in a
variety of relativistic nuclear structure models (mean field, Hartree and random
phase approximation (RPA), both in nuclear matter and finite nucleus), using a
simple reaction mechanism in which the distortion of the strongly interacting
kaons is accounted for through an effective number of nucleons participating in
the reaction, $N_{eff}$, i. e., 
\begin{equation}
  \frac{d^2\sigma}{d\Omega d\omega} = N_{eff} \frac{d\sigma_{K^{+}N}}{d\Omega}
    R(q,\omega) ,
\end{equation}
where $d\sigma_{K^{+}N}/d\Omega$ is the elementary cross section and 
$R(q,\omega)$ the nuclear response function. The agreement might look good, but
two observations are in order:
\begin{itemize}
\item[i)] In those calculations use has been made of the empirical
$d\sigma_{K^{+}N}/d\Omega$ and of the ``experimental'' $N_{eff}$, the latter
having been obtained by integrating the experimental QE cross sections
without accounting for any background. The values for $N_{eff}$ thus obtained
are $\sim30\%$ higher than Glauber theory predictions. Note that it is rather
difficult to interpret this increase of $N_{eff}$ through a modification of the
in-medium $K^{+}N$ cross section. A decrease of $\sigma_{K^{+}N}$ would give
rise to a larger $N_{eff}$, but should also reflect in a smaller
$d\sigma_{K^{+}N}/d\Omega$, making QE scattering little
sensitive to the in-medium $K^{+}N$ cross section \cite{Kor95}. 
Furthermore, a
decrease of $\sigma_{K^{+}N}$ would be at variance with the findings from 
elastic scattering.
\item[ii)] Even if one ascribes the larger $N_{eff}$ to some exotic effect, the
agreement of the relativistic models with the data turns out to be good at high
momenta (where RPA effects are smaller) and poor at low momenta (where
collectivity should be stronger). Although energy transfers below 
$\sim10\div15$ MeV should be discarded (since ther energy resolution of the 
experiment is not
sufficient to discriminate the elastic contamination), it is clear that the
strong distortion of the QE peak observed at low momenta is not
reproduced.
\end{itemize}

In Ref.\cite{DeP97} we have performed a calculation of $K^{+}$-nucleus
QE cross sections using a non-relativistic model for nuclear dynamics
and an implementation of Glauber theory up to two-step processes.
Details of the model can be found in Refs.\cite{DeP97,DeP93}.
Here, we briefly enumerate the main steps that have to be taken to get to the
QE cross section, starting with the nuclear response functions.
The latter are in general proportional to the imaginary part of the polarization 
propagator, which describes the propagation of density fluctuations in the
nuclear medium \cite{Fet71}:
\begin{itemize}
\item[-] The lowest order (uncorrelated) response is given by a mean field
described by a Woods-Saxon potential.
\item[-] An important class of single-particle correlations is embodied in the 
{\em spreading width} of the particle-hole (ph) states (i. e., the coupling of 
the ph states to higher order configurations). It can be accounted for by 
introducing a phenomenological complex ph self-energy.
\item[-] Although nuclear dynamics is non-relativistic, trivial relativistic
kinematical effects can be important at high transferred momenta, so that the
correct relativistic kinetic energies should be employed.
\item[-] Two-body correlations are introduced through a continuum RPA
calculation, using an effective ph interaction based on a $G$-matrix 
\cite{Nak84}. $G$-matrices are known to give rise to a too strong attraction
in the scalar-isoscalar channel. Although there exist many-body schemes that are
able to screen the $G$-matrix interaction, we have rather tried to see if
the data can put constraints on the effective interaction.
\end{itemize}
The reaction mechanism is based upon an implementation of the Glauber theory
up to two steps:
\begin{itemize}
\item[-] One-step contributions have been calculated by setting the coupling 
of kaons to the ph states according to the Glauber prescription, without 
resorting to the effective number approximation mentioned above.
The latter tends to overestimate collective effects, since it rescales in the
same way uncorrelated and correlated response functions, without accounting
for the fact that the nuclear excitations are generated in the low density
peripheral region of nuclei.
\item[-] The two-step contribution is much smoother (and, for kaons, turns 
out to be much 
smaller) than the one-step term and it can be safely calculated in the effective
number approximation, where it is proportional to the effective number of pairs
participating in the reaction and to the convolution of two QE responses.
\end{itemize}

\begin{figure}[t]
\begin{center}
\mbox{\epsfig{file=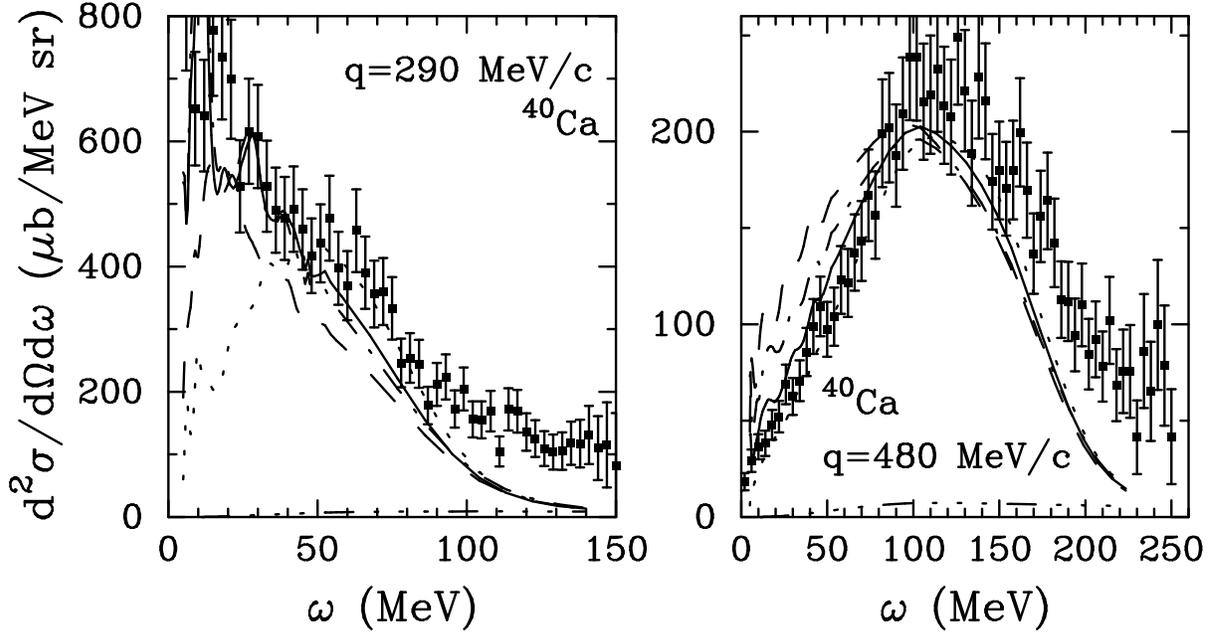,width=\textwidth}}
\caption{ QE cross sections for Ca: Free response (dot); RPA with 
full $G$-matrix (dash); RPA with the scalar-isoscalar interaction reduced 
by 50\% (solid); RPA with the renormalized interaction and the 
$N_{eff}$ approximation (dot-dash); two-step contribution (dot-dot-dot-dash).
}
\label{fig:ca40}
\end{center}
\end{figure}

Results are displayed in Figure \ref{fig:ca40} for $^{40}$Ca.
One can see that the strength and the shape of the responses are well
reproduced at all momenta. Collectivity manifests itself mainly on the left of
the QE peak: as mentioned above, the $G$-matrix gives too much attraction; the
data seem to point to a reduction of $\sim50\%$ of the effective interaction
in the scalar-isoscalar channel. Note also the smallness of the two-step term.
We also show results using the effective number 
approximation as in Refs.\cite{Kor93,Kor95}: as anticipated, there is some 
overestimation of collectivity, but the overall size of the response comes out 
correctly.
Moving along the high energy tail, the calculated cross sections tend to lay 
more and more below the data, suggesting the presence of a background (as it
happens, e. g., in (e,e') scattering), which would be responsible of the high
values for $N_{eff}$ found in Refs.\cite{Kor93,Kor95}.

Finally, note that use of the Glauber predictions for $N_{eff}$ in the
relativistic calculations of Figures \ref{fig:exp1} and \ref{fig:exp2} would 
result in underestimating the data by $\sim30\%$, essentially because of the 
RPA correlations of the relativistic models that quench the response functions,
while non-relativistic dynamical models give mainly rise to an enhancement at
low transferred energies.

\end{document}